\documentclass[10pt]{revtex4}
\usepackage{amssymb}
\usepackage{latexsym}
\usepackage{epsfig}
\begin{document}
\title{Thermodynamic of universe with a varying dark energy component}
\author{H. Ebadi$^{1,2}$\footnote{hosseinebadi@tabrizu.ac.ir}, H. Moradpour$^2$\footnote{h.moradpour@riaam.ac.ir}}
\address{$^1$ Astrophysics Department, Physics Faculty, University of Tabriz, Tabriz, Iran,\\
$^2$ Research Institute for Astronomy and Astrophysics of Maragha
(RIAAM), P.O. Box 55134-441, Maragha, Iran.}
\begin{abstract}
We consider a FRW universe filled by a dark energy candidate
together with other possible sources which may include the
baryonic and non-baryonic matters. Thereinafter, we consider a
situation in which the cosmos sectors do not interact with each
other. By applying the unified first law of thermodynamics on the
apparent horizon of the FRW universe, we show that the dark energy
candidate may modify the apparent horizon entropy and thus the
Bekenstein limit. Moreover, we generalize our study to the models
in which the cosmos sectors have a mutual interaction. Our final
result indicates that the mutual interaction between the cosmos
sectors may add an additional term to the apparent horizon entropy
leading to modify the Bekenstein limit. Relationships with
previous works have been addressed throughout the paper. Finally,
we investigate the validity of the second law of thermodynamics
and its generalized form in the interacting and non-interacting
cosmoses.
\end{abstract}

\maketitle

\section{Introduction}
Since the expanding universe is homogeneous and isotropic on
scales larger than about $100$-Mpc, it can be modeled by the
so-called FRW metric \cite{roos}
\begin{eqnarray}\label{frw}
ds^{2}=dt^{2}-a^{2}\left( t\right) \left[ \frac{dr^{2}}{1-kr^{2}}%
+r^{2}d\Omega ^{2}\right],
\end{eqnarray}
where $k=0,\pm1$ is the curvature constant corresponding to a
flat, closed and open universe, respectively. Additionally, $a(t)$
is the scale factor written as $a(t)=a_0t^{\frac{2}{3(1+\omega)}}$
for $\omega>-1$ and $a(t)=a_0\exp{Ht}$ when $\omega=-1$, whiles
$\omega=\frac{p}{\rho}$ is the state parameter of prefect
dominated fluid. In addition, $H\equiv\frac{\dot{a}}{a}$ is the
Hubble parameter \cite{roos}. Moreover, for Phantom regimes
($\omega<-1$) the scale factor is written as
$a(t)=a_0(t_{br}-t)^{\frac{2}{3(1+\omega)}}$, where $t_{br}$ is
the big rip singularity time, everything will be decomposed to its
fundamental constituents at that time, \cite{phan}. Additionally,
it is shown that one can use the conformal form of this metric to
describe the inhomogeneity of the cosmos in scales smaller than
$100$-Mpc \cite{rma}. In the standard cosmology a primary
inflationary expansion era is used to get a suitable theoretical
description for horizon problem which emerges in the study of
Cosmic Microwave Background (CMB) \cite{roos}. Observational data
signals us a universe with $\dot{a}\geq0$ and $\ddot{a}\geq0$
\cite{ac1,ac2,ac3,ac4}, which means that we need to modify the
gravitational theory \cite{mod,meeq,de} or considering an unknown
source, named dark energy (DE), for describing this phase of
expansion \cite{de,de1,mod1}. A simple model used to explain DE
considers an unknown fluid with constant density, pressure and
$\omega_D=-1$ called the cosmological constant (CC), and leads to
an exponential expansion ($a(t)=a_0\exp{Ht}$) \cite{roos}. We
should note that the current expanding phase of the universe is in
full agreement with both of the thermodynamics equilibrium
conditions and the rise of complexity content of the universe
meaning that the universe may maintain its current expanding phase
\cite{mr}. More studies on the thermodynamics of DE and the final
state of universe can be found in refs.~\cite{noj,noj1,pavon2}.
Bearing the primary inflationary era together with the CC model of
DE in mind, two difficulties including the \textit{fine tuning}
and \textit{coincidence} problems are inevitable \cite{roos}. It
is also useful to mention here that since the CC model has a
satisfactory match to the observational data, it formes a basis
for the standard cosmology \cite{roos}.

Observational data support a DE candidate with varying energy
density \cite{d1,d2,d3,d4,d5,d6}. Indeed, there are various
attempts to model the source of the primary and current
accelerating eras by introducing a varying model for the DE
candidate
\cite{meeq,de,de1,mod1,gde,ven,GGDE,GGDE1,sola,sola1,sola2,sola3,sola4,saha,tere1,tere,tere2,lima}.
Recently, Lima and co-workers proposed that a universe filled by a
dynamical vacuum energy density can avoid the big bang as well as
the big crunch singularities, the fine tuning and coincidence
problems \cite{lima}. Indeed, since the vacuum density is
decreased as a function of the Hubble parameter, the Lima's model
has enough potential for solving the fine tuning and coincidence
problems \cite{lima,lima2}. Additionally, because in their model,
the cosmos began to expand from a primary unstable de-Sitter
spacetime, and finally reaches to another eternal de-Sitter
spacetime, the horizon problem as well as the big crunch problem
are naturally solved \cite{lima}. Moreover, It is shown that the
ultimate de-Sitter spacetime is in accordance with the
thermodynamic equilibrium conditions, and therefore the cosmos may
serve its final stage \cite{lima1}. In this model, the state
parameter of the vacuum energy satisfies the $\omega_D=-1$
condition, and thus the vacuum energy decays into the other fields
confined to the apparent horizon of the FRW universe
\cite{lima,lima3}. It is worthwhile to mention here that the decay
of vacuum into the other fields is due to a mutual interaction
between the cosmos sectors leading to leave thermal fluctuations
into the cosmos in this model \cite{lima3}. It is a good feature,
because observations allows a mutual interaction between the
cosmos sectors \cite{int1,int11,ob1,ob2,ob3}. Thermodynamics of
such possible mutual interactions are also studied in various
theories of gravity by considering various models for DE
\cite{int11,te2,te3,te4,int3}. In fact, the relation between such
possible interactions, coincidence and fine tuning problems and
thermal fluctuations attracts more investigators to itself
\cite{lima3,int1,int2,int3}.

Similarity between the Black Holes laws and those of
thermodynamics motivates us to define a temperature as
\begin{eqnarray}\label{temp}
T=\frac{\kappa}{2\pi},
\end{eqnarray}
where $\kappa$ is the surface gravity of Black Hole \cite{pois}.
On one hand, for some spacetimes, such as the de-Sitter spacetime,
surface gravity and thus the corresponding temperature are
negative \cite{pois}, and therefore we need to define
$T=\frac{|\kappa|}{2\pi}$ in order to get the positive values for
temperature \cite{CaiKim,GSL1}. Whiles, on the other hand, one can
get the Einstein equations on the event horizon of Black Holes (as
a causal boundary) by applying the first law of thermodynamics on
the event horizon and considering Eq.~(\ref{temp}) as a suitable
definition for temperature \cite{J1,T11,J11}. Indeed, it seems
that this similarity is much more than a mere resemblance
\cite{J1,J11,M1,M11,T1,T11,T12,T13,T14,T15,r1,r2,r3}.

The apparent horizon of the FRW universe, as the marginally
trapped surface, is evaluated by
\begin{eqnarray}\label{ah2}
\partial_{\alpha}\zeta\partial^{\alpha}\zeta=0\rightarrow r_H,
\end{eqnarray}
where $\zeta=a(t)r$, and can be considered as the causal boundary
\cite{Hay2,Hay22,Bak}. Therefore, One gets \cite{sheyw1,sheyw2}
\begin{eqnarray}\label{ah}
\tilde{r}_A=\frac{1}{\sqrt{H^2+\frac{k}{a(t)^2}}}.
\end{eqnarray}
Moreover, the surface gravity associated with the apparent horizon
of the FRW universe can be evaluated by using
\begin{equation}\label{SG}
\kappa=\frac{1}{2\sqrt{-h}}\partial_{a}(\sqrt{-h}h^{ab}\partial_{b}\zeta).
\end{equation}
where $h_{ab}=\textmd{diag}(-1,a(t)^2)$ \cite{sheyw1,sheyw2}.
Since the WMAP data indicates a flat universe, from now we set
$k=0$ \cite{roos,phan}. Thus, simple calculations lead to
\begin{eqnarray}
\kappa=-H(1+\frac{\dot{H}}{2H^2}),
\end{eqnarray}
and therefore
\begin{eqnarray}\label{t}
T=\frac{\kappa}{2\pi}=-\frac{H}{2\pi}(1+\frac{\dot{H}}{2H^2}).
\end{eqnarray}
where we have used Eq.~(\ref{temp}) to obtain this equation
\cite{GSL1,Bak,Cai3,hel,hel1}. It is useful to note here that for
the FRW universe supported by a fluid with $\rho=-p=constant$
($\omega=-1$), $\dot{H}=0$ and therefore this equation covers the
result of de-Sitter spacetime ($T=-\frac{H}{2\pi}$)
\cite{pois,GSL1}.

In cosmological setups, some authors use various definition of
temperature and get the corresponding Einstein equations
(Friedmann equations) on the apparent horizon
\cite{Cai2,Cai3,CaiKim,GSL1,Bak,hel,hel1}. In order to avoid the
negative temperature, authors in \cite{CaiKim}, have defined
$T=\frac{H}{2\pi}\simeq\frac{|\kappa|}{2\pi}$ and used the first
law of thermodynamics (in the $TdS_A=-dQ$ form) to get the
Friedmann equations. In their approach $S_A=\pi\tilde{r}_A^2$ (the
Bekenstein limit) and $Q$ are the horizon entropy and energy
crossed the apparent horizon, respectively \cite{GSL1}. Indeed,
authors argued that the extra minus sign in the first law of
thermodynamics is the result of universe expansion leading to
decrease the energy of confined fluid together with increase the
size of the universe and thus $S_A$. Therefore, by using original
definition of temperature~(\ref{temp}) and thus~(\ref{t}), called
the Hayward-Kodama temperature \cite{GSL1,Bak,hel,hel1}, together
with the $TdS_A=dQ$ form of the first law of thermodynamics we can
cover the Friedmann equation. Moreover, it seems that
$W=\frac{1}{2}h_{ab}T^{ab}$, where $T^{ab}$ is the energy momentum
tensor of fluid which spreads over the cosmos, plays the role of
pressure in the dynamics spacetimes and thus the FRW universe
\cite{Cai2,Cai3,sh1,GSL1,Bak,hel,hel1}. Following this argument,
authors in \cite{Cai2,Cai3,Bak,hel,hel1,GSL1} have
used~(\ref{temp}) and the work density definition ($W$) to get the
Friedmann equations by applying the first law of thermodynamics
($TdS_A=dQ=dE-WdV$) on the apparent horizon of the FRW universe in
various theories of gravity, whiles $E$ is the energy confined to
the apparent horizon. It is also shown that Loop Quantum Gravity
corrects the horizon entropy which leads to modify the Friedmann
equations on the apparent horizon if one considers~(\ref{temp})
together with $TdS_A=dE-WdV$ \cite{sh1,r1}. The entropy of a
self-gravitating system depends on the gravitational theory used
to describe the gravity field. Accordingly, it seems that the
self-gravitating systems satisfy the Bekenstein limit of entropy
in the Einstein general relativity framework. But, since the
origin of DE is unknown, it may have either a geometrical or
physical origin, one can expect that the DE candidate may affect
the horizon entropy. By the same token, it is shown that the ghost
dark energy and its generalization, as the dynamics candidates for
DE, may also add an additional term to the entropy of various
horizons leading to modify the Bekenstein limit \cite{cana,cana1}.
Therefore, it seems that the dynamics model of DE may lead to
modify the horizon entropy and thus, the Bekenstein limit.
Recently, it is also shown that a mutual interaction between the
cosmos sectors may change the horizon entropy \cite{mitra}. The
second law of thermodynamics states that the horizon entropy may
meet the $\frac{dS_A}{dt}\geq0$ condition \cite{haw}. Nowadays,
thanks to the Bekenstein works \cite{bek,bek2}, it is believed
that the rate of the total entropy of a gravitational system
should be positive meaning that
$\frac{dS_A}{dt}+\frac{dS_{in}}{dt}\geq0$, while $S_{in}$ is the
entropy of confined fluid. The latter is called the general second
law of thermodynamics \cite{bek,bek2,GSL1}. Comprehensive reviews
on the various temperature definitions in cosmological setups,
their motivations together with the validity of the first, second
and generalized second laws of thermodynamics can be found in
refs.~\cite{hel,hel1,GSL1}. Now, one can ask how a DE candidate
and its probable interaction with other parts of cosmos affect the
horizon entropy, the second and generalized second laws of
thermodynamics?

In this paper, we point to the unified first law of thermodynamics
and assume that it is available on the apparent horizon of the
flat FRW universe, while $T$ (the horizon entropy) corresponds to
the Hayward-Kodama definition of temperature~(\ref{t}) on the
apparent horizon of the FRW universe~\cite{Bak,hel,hel1}, and show
that a DE candidate may lead to a new bound for the horizon
entropy, whiles the cosmos sectors do not interact with each
other. Additionally, we show that any mutual interaction between
the cosmos sectors may also modify the horizon entropy. The
relationships with similar works are also studied. Moreover, the
results of considering the Cai-Kim temperature are also derived.
Finally, the validity of the second law of thermodynamics and its
generalization is also addressed. Since the physics behind the
Lima's model~\cite{lima} is completely different from the ordinary
models, introducing for describing DE, we point to results of
considering this model.

the paper is organized as follows. In the next section, after a
brief review on the previous related works, we apply the unified
first law of thermodynamics on the apparent horizon of the flat
FRW universe, and show that how a dynamic candidate for DE may
change the horizon entropy, whiles the cosmos sectors do not
interact with each other. Thereinafter, we generalize our study to
the interacting case and get a modification for the horizon
entropy due to the mutual interaction between the cosmos sectors.
In section ($\textmd{III}$), we study the validity of the second
law of thermodynamics and its generalization. Section
($\textmd{IV}$) is devoted to a summary and concluding remarks.
Throughout this paper, we set $G=c=\hbar=1$ for simplicity.
\section{Horizon Entropy and the unified first law of thermodynamics}
The unified first law of thermodynamics, which is available in
some theories of gravity, is written as
\begin{eqnarray}\label{ufl}
dE=A\Psi+WdV,
\end{eqnarray}
where $W=\frac{1}{2}h_{ab}T^{ab}$ and
$E=\frac{\zeta}{2}(1-h^{ab}\partial_a \zeta \partial_b
\zeta)|_{\zeta=\tilde{r}_A}$ are the work density and the
Misner-Sharp energy confined to the apparent horizon, respectively
\cite{CaiKim,Hay2,Bak,r1,r2,r3,cana,mitra,cana1,Hay22}. In
addition, $A$ and $\Psi$ are the area of horizon and the energy
supply vector, respectively, and
\begin{eqnarray}\label{ufl1}
A\Psi=A\psi_a dx^a,
\end{eqnarray}
while
\begin{eqnarray}\label{ufl2}
\psi_a = T^b_a\partial_b \zeta + W\partial_a \zeta,
\end{eqnarray}
is the projection of the total four-dimensional energy-momentum
tensor $T_{\mu \nu}$ in the normal direction of the
two-dimensional sphere. Consider a perfect fluid source
($T^{\mu}_{\nu}=diag(-\rho_T,p_T,p_T,p_T)$) together with
Friedmann equations, by simple calculations we get $E=\rho_T V$,
\begin{eqnarray}\label{uf3}
dE-WdV=Vd\rho_T+\frac{p_T+\rho_T}{2}dV,
\end{eqnarray}
and
\begin{eqnarray}\label{uf4}
A\Psi=-AH\zeta(\frac{\rho_T+p_T}{2})dt+Aa(\frac{\rho_T+p_T}{2})dr,
\end{eqnarray}
where $a$ is the scale factor. Using the energy-momentum
conservation law ($\nabla^{\mu}T_{\mu \nu}=0$)
\begin{eqnarray}\label{energymomentum0}
\dot{\rho}_T+3H(\rho_T+p_T)=0,
\end{eqnarray}
and $adr=d\zeta-rda$ in rewriting Eq.~(\ref{uf4}) to obtain
\begin{eqnarray}\label{uf5}
A\Psi=Vd\rho_T+\frac{p_T+\rho_T}{2}dV,
\end{eqnarray}
where $dV=Ad\zeta$ and $A=4\pi\zeta^2$. By comparing this equation
with~(\ref{uf3}), we get
\begin{eqnarray}\label{uf6}
A\Psi=dE-WdV,
\end{eqnarray}
which is the unified first law of thermodynamics. This result is
independent of the number and nature of fluids which support the
background spacetime. In addition, one may decompose $T_{\mu \nu}$
into
\begin{eqnarray}\label{ef7}
T_{\mu \nu}=T_{\mu \nu}^{DE}+T_{\mu \nu}^{m},
\end{eqnarray}
where $T_{\mu \nu}^{DE}$ and $T_{\mu \nu}^{m}$ are the energy
momentum tensors of DE and the material parts of cosmos
(radiation, matter and etc.), respectively. In this situation, it
is apparent that $\Psi=\Psi^{DE}+\Psi^{m}$, $W=W^{DE}+W^{m}$ and
$E=E^{DE}+E^{m}$, where $E^{DE}=\rho^{DE}V$ and $E^{m}=\rho^{m}V$.
Therefore, by following the above argument, whenever
$\nabla^{\mu}T_{\mu \nu}^{m}=\nabla^{\mu}T_{\mu \nu}^{DE}=0$, we
get
\begin{eqnarray}\label{uf81}
A\Psi^m= A\Psi-A\Psi^{DE}=dE^{m}-W^{m}dV.
\end{eqnarray}
$\delta Q$ (the heat flow crossing the horizon) is determined by
the pure matter energy-momentum tensor ($T^m_{\mu \nu}$) as
\cite{r1,r2,r3,CaiKim,cana,cana1,mitra}
\begin{eqnarray}\label{ufl100}
\delta Q\equiv A\Psi^m.
\end{eqnarray}
For some gravitational theories, one can use the Clausius relation
together with Eq.~(\ref{t}) to get the horizon entropy ($S_A$) by
using \cite{r1,r2,r3,CaiKim,cana,cana1,mitra}
\begin{eqnarray}\label{ufl10}
TdS_A=\delta Q\equiv A\Psi^m,
\end{eqnarray}
which leads to
\begin{eqnarray}\label{uf8}
TdS_A=A\Psi^m= A\Psi-A\Psi^{DE}=dE^{m}-W^{m}dV,
\end{eqnarray}
where we have used~(\ref{uf81}) to get the last equation. It is
useful to note that for some theories such as the $f(R)$ gravity,
Eq.~(\ref{ufl10}) and thus~(\ref{uf8}) is not always available
\cite{r2}.

Recently, some authors considered the Hayward-Kodama definition of
temperature~(\ref{t}), a universe filled by either a ghost dark
energy or its generalized form together with a pressureless matter
and use the $TdS_A=A\Psi-A\Psi^{DE}$ relation to get an expression
for the entropy ($S_A$) \cite{cana}. Their results show that the
entropy of the matter fields differs from the Bekenstein entropy
due to the DE effects. They argued that their results are in
agreement with the entropy of the apparent horizon in the DGP
braneworld model which signals that this approach may be used to
get the entropy of apparent horizon in other theories of gravity.
This adaptation between these entropies signals that one may find
a geometrical interpretation for the origin of the ghost dark
energy model (as a DE candidate) by using the DGP braneworld model
of gravity. Motivated by this work, Sheykhi extended their work to
the apparent horizon of the FRW universe and used the
$TdS_A=dE^{m}-W^{m}dV$ relation to get the same result as that of
ref.~\cite{cana} for the entropy. Finally, he concludes that the
obtained relation for the entropy ($S_A$) may be interpreted as
the corrected relation for the apparent horizon entropy
\cite{cana1}. It is useful to stress here that Eq.~(\ref{uf81})
clarifies the reason of getting the same result for the horizon
entropy by authors in Refs.~\cite{cana,cana1}. Moreover, their
results are available only when $\nabla^{\mu}T_{\mu
\nu}^{m}=\nabla^{\mu}T_{\mu \nu}^{DE}=0$ which means that the
cosmos sectors do not interact with each other. Recently, by
considering the FRW universe in which the cosmos sectors interact
with each other, Mitra et al. use the $TdS_A=A\Psi-A\Psi^{DE}$
relation to get the trapping horizon entropy in the Einstein
relativity framework. They argued that the obtained relation for
the entropy differs from the Bekenstein entropy due to the mutual
interaction between the cosmos sectors \cite{mitra}. In continue,
Mitra et al. extended their hypothesis to other different gravity
theories \cite{mitra}. Moreover, it is also shown that a
gravitationally induced particle production process as the DE
candidate may change the horizon entropy \cite{lima3}. Bearing the
Lovelock theory in mind, some authors used the
$TdS_A=A\Psi-A\Psi^{DE}$ relation to get the entropy of the
apparent horizon in cosmological setup \cite{r3}. Another study
including the loop quantum cosmology can be found in ref
\cite{r1}. Here, by considering a varying DE candidate, we are
going to find a general relation for the entropy of the apparent
horizon in both of the interacting and non-interacting cosmoses
and investigate the second and generalized second laws of
thermodynamics in the Einstein relativity frame work where
Eq.~(\ref{ufl10}) and thus~(\ref{uf8}) are valid.

For this propose, consider the flat FRW universe with Friedmann
equation
\begin{eqnarray}\label{fried1}
H^2=\frac{8\pi}{3}(\rho+\rho_D),
\end{eqnarray}
where $\rho_D$ is the density of dark energy component. In
addition, $\rho$ is the density of rest fluids in the cosmos which
may include baryonic matters, dark matter and etc, leading to
\begin{eqnarray}\label{oden}
\rho=\rho_{bm}+\rho_{DM}+...
\end{eqnarray}
Therefore, $\rho$ is nothing but $\rho^m$ which is previously
introduced. For the sake of simplicity, we omit the $m$ label
throughout the paper. From Eq.~(\ref{fried1}) and the Bianchi
identity we get
\begin{eqnarray}\label{friedde}
2HdH-\frac{8\pi}{3}d\rho_D=\frac{8\pi}{3}d\rho,
\end{eqnarray}
and
\begin{eqnarray}\label{energymomentum}
\dot{\rho}+3H(\rho+p)+\dot{\rho}_{D}+3H(\rho_{D}+p_{D})=0,
\end{eqnarray}
which is nothing but the energy-momentum conservation law,
respectively. In this equation, $p_D$ and $p$ denote the vacuum
pressure and the pressure corresponding to the density $\rho$,
respectively. Dot is also denoted as the time derivative. Consider
a dark energy candidate with density profile
\begin{eqnarray}\label{dden}
\rho_D=\frac{3\alpha+3\beta H^2+3\gamma H^{2n}}{8\pi},
\end{eqnarray}
which converges to CC whiles $\beta=\gamma=0$. Whiles
$n=\frac{1}{2}$, it covers the ghost dark energy model and its
generalization for $\alpha=\beta=0$ and $\alpha=0$, respectively
\cite{gde,ven,GGDE}. The $\gamma=0$ case has been extensively
studied in the literatures
\cite{GGDE1,sola,sola1,sola2,sola3,sola4}. The results of
considering either an arbitrary value for $n$ or optional function
of $H$ for $\rho_D=f(H)$ can be found in \cite{saha}.

Moreover, the cosmological applications of considering model with
$\alpha=0$, $n=\frac{3}{2}$ and $\omega_D=-1$  has also been
studied \cite{tere1,tere}. More similar density profiles for the
DE candidate with $\omega_D=-1$ can also be found in \cite{tere2}.
Another attractive case proposed by Lima et al. is obtainable by
imposing the $n>1$ condition together with $\omega_D=-1$ to the
density profile of the DE candidate, whenever $n$ is also an
integer number \cite{lima}. It is useful to note that $E_D=\rho_D
V$ and $E=\rho V$ are the energy of dark energy component and the
energy corresponding to the density $\rho$, respectively.
Therefore, $E$ is nothing but $E^m$ mentioned previously and we
omit the $m$ label for the sake for simplicity.
\subsection{Non-Interacting Models}
At the first step we consider a universe in which the cosmos
sectors do not interact with each other. Therefore, the
energy-momentum conservation law implies~(\ref{energymomentum})
\begin{eqnarray}\label{ocon1}
\dot{\rho}+3H(\rho+p)=0,
\end{eqnarray}
and
\begin{eqnarray}\label{dcon1}
\dot{\rho}_{D}+3H(\rho_{D}+p_{D})=0.
\end{eqnarray}
Substituting~(\ref{dden}) into~(\ref{fried1}) to get
\begin{eqnarray}\label{oden1}
\frac{8\pi}{3}\rho=H^2-\alpha-\beta H^2-\gamma H^{2n}.
\end{eqnarray}
Bearing Eq.~(\ref{ocon1}) in mind, by using Eq.~(\ref{friedde}) we
reach at
\begin{eqnarray}\label{dif1}
(2H(1-\beta)-2n\gamma H^{2n-1})dH=-8\pi H(\rho+p)dt.
\end{eqnarray}
Using the Hayward-Kodama temperature relation
($-T=\frac{H}{2\pi}(1+\frac{\dot{H}}{2H^2})$)~\cite{GSL1} to
obtain
\begin{eqnarray}\label{dif2}
-T(2H(1-\beta)-2n\gamma H^{2n-1})dH=-4H^2(\rho+p)dt-2(\rho+p)dH.
\end{eqnarray}
From Eq.~(\ref{ocon1}), since $E=\rho V$ and
$dV=-\frac{4\pi}{H^4}dH$, we get $dE=-4\pi \rho H^{-4}dH-4\pi
H^{-2}(\rho+p)dt$ leading to
\begin{eqnarray}\label{dif3}
(\rho+p)dt=-\frac{H^2dE}{4\pi}-\frac{\rho dH}{H^2}.
\end{eqnarray}
If we combine this equation with~(\ref{dif2}) we obtain
\begin{eqnarray}\label{dif4}
T(-2H(1-\beta)+2n\gamma H^{2n-1})dH=\frac{H^4}{\pi}dE+2(\rho-p)dH.
\end{eqnarray}
It is easy to show that this equation can be rewritten as
\begin{eqnarray}\label{dif5}
T[(-\frac{2\pi}{H^3}(1-\beta)+2n\gamma\pi H^{2n-5})dH]=dE-WdV.
\end{eqnarray}
In this equation $W=\frac{\rho-p}{2}$ is the work density required
for applying a hypothetical displacement $d\tilde{r}_A$ to the
apparent horizon \cite{cana1,mitra}. By comparing this result with
Eq.~(\ref{uf8}), one gets
\begin{eqnarray}\label{hentropy1}
dS_A=(-\frac{2\pi}{H^3}(1-\beta)+2n\gamma\pi H^{2n-5})dH
\end{eqnarray}
leading to
\begin{eqnarray}\label{hentropy}
S_A=\frac{A}{4}(1-\beta)+\frac{n\gamma\pi^{n-1}}{n-2}A^{2-n},
\end{eqnarray}
where $A=4\pi\tilde{r}_A^2=\frac{4\pi}{H^2}$ is the area of
horizon. Therefore, $\frac{n\gamma\pi^{n-1}}{n-2}A^{2-n}$ is a new
term besides the area term. In addition, since the entropy is not
an absolute quantity, we have set the integral constant to zero.
It is also apparent that, for $n=2$, entropy is not well-defined.
In order to eliminate this weakness, let us restart from
Eq.~(\ref{hentropy1}), by substituting $n=2$ and taking
integration from that, we get
\begin{eqnarray}\label{hentropy2}
S_A=\frac{A}{4}(1-\beta)-\frac{\gamma \pi \ln\pi}{2}+\frac{\gamma
\pi \ln A}{2}+S_0.
\end{eqnarray}
Finally, since entropy is not an absolute quantity, one can set
$S_0=\frac{\gamma \pi \ln\pi}{2}$, and gets
\begin{eqnarray}\label{hentropy3}
S_A=\frac{A}{4}(1-\beta)+\frac{\gamma \pi \ln A}{2}.
\end{eqnarray}
Therefore, models with $n=2$ induce a logarithmic correction to
the horizon entropy. Logarithmic correction terms have been
previously proposed by some authors which either consider the
thermal equilibrium and quantum fluctuations in loop quantum
gravity framework \cite{l1,l2,l3,l4,l5,l6,l7,l8,l9,l,l0,l10} or
the thermal fluctuations of system about its thermodynamic
equilibrium state \cite{l11,l12}. Indeed, logarithmic correction
due to the thermal fluctuations are valid in all physical systems
\cite{lan}. Let us study some choices with $n=\frac{1}{2}$.
Bearing Eq.~(\ref{hentropy}) in mind, For a constant vacuum energy
density ($\rho_D=\alpha$), we face with the $\Lambda CDM$ theory
and we get $S_A=\frac{A}{4}$ which is in agreement with previous
studies \cite{Cai2,Cai3,CaiKim}. Moreover, for $\alpha=0$,
$\beta=0$ and $n=\frac{1}{2}$, we have
\begin{eqnarray}\label{gde}
\rho_D=\frac{3\gamma}{8\pi}H,
\end{eqnarray}
which is the profile density of ghost dark energy model
\cite{ven,gde}. In this limit, from Eq.~(\ref{hentropy}), we get
\begin{eqnarray}\label{hentropygde}
S_A=\frac{A}{4}-\frac{\gamma}{3\sqrt{\pi}}A^{\frac{3}{2}},
\end{eqnarray}
which is in agreement with the ghost dark energy modification to
the entropy evaluated previously \cite{cana,cana1}. Here, we have
used the Hayward-Kodama definition of temperature~(\ref{t})
together with the apparent horizon of the FRW universe to get this
relation whiles, author in \cite{cana1}, has considered
$T=\frac{|\kappa|}{2\pi}$ to get~(\ref{hentropygde}) on the
apparent horizon. Moreover, authors in~\cite{cana} used trapping
horizon and the temperature definition $T=\frac{|\kappa|}{2\pi}$
to get this relation. Additionally, equation~(\ref{dden}), for
$\alpha=0$ and $n=\frac{1}{2}$, reduces to
\begin{eqnarray}\label{ggde}
\rho_D=\frac{3\beta}{8\pi}H^2+\frac{3\gamma}{8\pi}H,
\end{eqnarray}
which is the profile density of generalized ghost dark energy
model \cite{GGDE,ggde2}. By considering this profile density we
get
\begin{eqnarray}\label{hentropyggde}
S_A=\frac{A}{4}(1-\beta)-\frac{\gamma}{3\sqrt{\pi}}A^{\frac{3}{2}},
\end{eqnarray}
as the modification of the generalized ghost dark energy model to
the horizon entropy \cite{cana}. Although this result is
previously obtained by authors in ref.~\cite{cana}, but our
derivation is completely different from that of they. Here, we
worked on the apparent horizon whiles they have considered the
trapping horizon and found the similar results. In a more general
case, for arbitrary functional form of $\rho_D$, by following the
above recipe we get
\begin{eqnarray}\label{general}
dS_A=(-\frac{2\pi}{H^3}+\frac{8\pi^2}{3H^4}\rho_D^{\prime})dH,
\end{eqnarray}
where prime denotes derivative with respect to $H$. Taking
integral to obtain
\begin{eqnarray}\label{general1}
S_A=\frac{A}{4}+\frac{8\pi^2}{3}\int\frac{1}{H^4}d\rho_D +C,
\end{eqnarray}
where $C$ is the integral constant. Therefore, a varying DE
candidate imposes a correction term to the horizon entropy in
accordance with the first law of thermodynamics and thus, the
second term of the RHS of Eq.~(\ref{general1}). It is also useful
to note that the result of considering CC ($S_A=\frac{A}{4}$) is
obtainable by substituting $d\rho_D=0$ in this equation
\cite{Cai2,Cai3}.

Now, let us use the Cai-Kim temperature ($T=\frac{H}{2\pi}$)
\cite{CaiKim} to get the entropy of apparent horizon. In order to
achieve this goal, we follow the approach of authors in
ref.~\cite{CaiKim}, where $TdS_A=-dQ$ and $dV=0$. Using this
argument and bearing Eqs.~(\ref{ufl10}) and~(\ref{uf8}) in mind to
reach
\begin{eqnarray}
dS_A=-\frac{V}{T}d\rho.
\end{eqnarray}
Now, by substituting $d\rho$ from Eq.~(\ref{friedde}) into this
equation, one gets
\begin{eqnarray}
dS_A=-\frac{2\pi}{H^3}dH+\frac{8\pi^2}{3H^4}d\rho_D,
\end{eqnarray}
which leads to
\begin{eqnarray}
S_A=\frac{A}{4}+\frac{8\pi^2}{3}\int \frac{d\rho_D}{H^4}+C,
\end{eqnarray}
where $C$ is the integration constant. Therefore, once again, we
get a relation for the horizon entropy which is in full agreement
with the previous result~(\ref{general1}), obtained by considering
the Hayward-Kodama temperature.
\subsection{Interacting Models}
When the cosmos sectors interact with each other, energy-momentum
conservation law implies~(\ref{energymomentum})
\begin{eqnarray}\label{energymomentum12}
\dot{\rho}+3H(\rho+p)=-\dot{\rho}_{D}-3H(\rho_{D}+p_{D}),
\end{eqnarray}
meaning that
\begin{eqnarray}\label{energymomentum12f}
d\rho=-3H(\rho+p)dt-d\rho_{D}-3H(\rho_{D}+p_{D})dt.
\end{eqnarray}
Therefore, by considering Eq.~(\ref{friedde}) and following the
recipe which leads to Eq.~(\ref{general1}), we get
\begin{eqnarray}\label{general2}
dS_A=-\frac{2\pi}{H^3}dH-\frac{8\pi^2}{H^3}(\rho_D+p_D)dt,
\end{eqnarray}
which yields
\begin{eqnarray}\label{general12}
S_A=\frac{A}{4}-8\pi^2\int\frac{\rho_D+p_D}{H^3}dt + C,
\end{eqnarray}
where $C$ is again an integral constant. Therefore, the second
term of RHS of this equation is nothing but the entropy correction
due to the mutual interaction between the cosmos sectors. For
interacting models in which the state parameter of the DE
candidate meets the $\omega_D=-1$ condition, and therefore
$\rho_D+p_D=0$, this additional term is zero meaning that the
horizon entropy in these models satisfies the Bekenstein limit
\cite{bek}. For instance, in the model proposed by Lima et al.
\cite{lima}, in which vacuum decays into the other parts of cosmos
and $\rho_D+p_D=0$, the horizon entropy of the flat FRW universe
meets the Bekenstein limit \cite{bek}. It is in agreement with the
initial and final de-Sitter spacetimes of this model, since the
horizon of de-Sitter spacetime meets the $S_A=\frac{A}{4}$
condition \cite{Cai2,Cai3,CaiKim}. Now, let us derive
Eq.~(\ref{general12}) by using the unified first law of
thermodynamics. Bearing the definition of $\Psi$ in mind, simple
calculations lead to
\begin{eqnarray}\label{51}
A\Psi^{DE}=-\frac{3V(\rho_D+p_D)H}{2}dt+\frac{A(\rho_D+p_D)}{2}[d\zeta-\zeta
H dt],
\end{eqnarray}
where we have used the $rda=\zeta H dt$ relation to obtain this
equation. It is a matter of calculation to show
\begin{eqnarray}
A\Psi^{DE}=-\frac{4\pi(\rho_D+p_D)}{H^2}[1+\frac{\dot{H}}{2H^2}]dt,
\end{eqnarray}
where we have used $dV=-\frac{3V}{H}\dot{H}dt$ to get this
equation. Since we work in the Einstein general relativity
framework, Eq.~(\ref{ufl10}) is valid, and thus, simple
calculations lead to
\begin{eqnarray}\label{53}
TdS_A=A\Psi-A\Psi^{DE}=-\frac{H}{2\pi}[1+\frac{\dot{H}}{2H^2}](-\frac{2\pi}{H^3}dH-\frac{8\pi^2}{H^3}(\rho_D+p_D)dt),
\end{eqnarray}
where we have used the $A\Psi=T(-\frac{2\pi}{H^3}dH)$ relation,
while $T$ is the Hayward-Kodama temperature, in obtaining this
relation \cite{GSL1,Bak,hel,hel1,mitra}. It is apparent that this
equation is nothing but~(\ref{general2}) which leads to
Eq.~(\ref{general12}).

Our result is in agreement with the recent work by Mitra et al.
\cite{mitra}. Whereas, we have started from the Friedmann
equations and considered the apparent horizon as the causal bound,
Mitra et al. used the trapping horizon and relation $\delta
Q^{m}\equiv A\Psi-A\Psi^{DE}$ to obtain~(\ref{general12}). It is
apparent that Eq.~(\ref{53}) clarifies that why both of us get the
same results, while, our start points differ from each other.

Finally, let us consider the Cai-Kim temperature to estimate the
horizon entropy. In this situation, for an infinitesimal time
$dV=0$, and from Eqs.~(\ref{uf5}) and~(\ref{51}) we get
\begin{eqnarray}\label{530}
TdS_A=-A\Psi^m=-A\Psi+A\Psi^{DE}=-V(d\rho+d\rho_D)-\frac{4\pi(\rho_D+p_D)}{H^2}dt,
\end{eqnarray}
where we have followed the approach of authors in
ref.~\cite{CaiKim} in order to define $TdS_A=-\delta Q$. Now,
bearing Eq.~(\ref{fried1}) in mind, since $T=\frac{H}{2\pi}$,
simple calculations lead to
\begin{eqnarray}\label{533}
S_A=\frac{A}{4}-8\pi^2\int\frac{\rho_D+p_D}{H^3}dt + C,
\end{eqnarray}
where $C$ is an integration constant. Therefore, by using the
Cai-Kim temperature and taking into account an infinitesimal time,
we get the same result for the horizon entropy as the result
obtained in Eq.~(\ref{general12}).
\section{The second and generalized second laws of thermodynamics}
On one hand, since cosmos is enclosed by the apparent horizon, it
forms a closed system and therefore, the entropy of its horizon
should increase during the universe expansion meaning that
\cite{haw}
\begin{eqnarray}\label{SL}
\frac{dS_A}{dt}\geq0.
\end{eqnarray}
It is called the second law of thermodynamics. Whereas, on the
other hand, the total entropy of the closed systems should be
increased. Since cosmos includes spacetime and its contents, which
includes the fluids supporting the geometry of background
spacetime, its total entropy consists of two parts including the
horizon ($S_A$) and the confined fluids components ($S_{in}$)
\cite{bek,bek2}. In fact, the generalized second law of
thermodynamics states that the rate of the total entropy of cosmos
including the horizon and confined fluids entropies cannot be
negative or briefly \cite{bek,bek2}
\begin{eqnarray}\label{GSL}
\frac{dS_A}{dt}+\frac{dS_{in}}{dt}\geq0.
\end{eqnarray}
Indeed, the total entropy of gravitational systems should
meet~(\ref{GSL}) \cite{bek,bek2}. But, here we point to the
required conditions for satisfying both of the above criterions.
\subsection{Non-Interacting case}
For the non-interacting cases and while $\rho_D$
meets~(\ref{dden}), by taking a time derivative of the Friedmann
equation~(\ref{fried1}) and using the energy-momentum conservation
law~(\ref{ocon1}) to get the Raychaudhuri equation
\begin{eqnarray}\label{rey}
\dot{H}=-4\pi(\rho+p)\frac{1}{1-\beta-n\gamma H^{2n-2}}.
\end{eqnarray}
Since during the cosmos life $\dot{H}<0$ \cite{roos}, we get
$1-\beta-n\gamma H^{2n-2}>0$ leading to
$H<(\frac{1-\beta}{n\gamma})^{\frac{1}{2n-2}}$ while $\rho+p>0$,
and $1-\beta-n\gamma H^{2n-2}<0$ which yields
$H>(\frac{1-\beta}{n\gamma})^{\frac{1}{2n-2}}$ for $\rho+p<0$.
Using Eqs.~(\ref{dif2}) and~(\ref{hentropy1}) to obtain
\begin{eqnarray}\label{hentropyf}
T\frac{dS_A}{dt}=-\frac{4\pi(\rho+p)}{H^2}[1+\frac{\dot{H}}{2H^2}].
\end{eqnarray}
It seems that horizons may satisfy the second law of
thermodynamics meaning that the $dS_A\geq0$ condition should be
valid \cite{pois,Cai2,Cai3,CaiKim}. In order to check the validity
of the second law of thermodynamics we insert
$T=-\frac{H}{2\pi}(1+\frac{\dot{H}}{2H^2})$ into this equation,
and get
\begin{eqnarray}\label{hentropyf2}
\frac{dS_A}{dt}=\frac{8\pi^2(\rho+p)}{H^3},
\end{eqnarray}
meaning that the second law of thermodynamics is available for the
apparent horizon whiles $\rho+p\geq0$. This conditions leads to
$\omega\geq-1$ for the state parameter $\omega$. Moreover, by
combining Eqs.~(\ref{hentropy}) and~(\ref{dcon1}) with together,
we get
\begin{eqnarray}\label{fff}
\frac{dS_A}{dt}=-\frac{2\pi\dot{H}}{H^3}(1+4\pi\frac{\rho_D+p_D}{\dot{H}}),
\end{eqnarray}
which means that the second law of thermodynamics is satisfied if
$1+4\pi\frac{\rho_D+p_D}{\dot{H}}\geq0$. Finally, the second law
of thermodynamics ($\frac{dS_A}{dt}\geq0$) is met by the horizon
component when $\rho_D+p_D\geq-\frac{\dot{H}}{4\pi}$ and
$\rho+p\geq0$ are satisfied, simultaneously. It is useful to
mention here that one can get
\begin{eqnarray}\label{rey0}
\dot{H}=-4\pi(\rho+p+\rho_D+p_D),
\end{eqnarray}
by equating Eqs.~(\ref{fff}) and~(\ref{hentropyf2}), which is
nothing but the Raychaudhuri equation obtainable by taking time
derivative from Eq.~(\ref{fried1}) and
using~(\ref{energymomentum}). Therefore, when $\rho+p\geq0$ and
$\frac{\dot{H}}{4\pi}\geq-(\rho_D+p_D)$ are available, then
$\rho+p+\frac{\dot{H}}{4\pi}\geq-(\rho_D+p_D)$ is obtainable which
is in agreement with the Raychaudhuri equation~(\ref{rey0}). For
the fluids confined to the apparent horizon with total density
$\rho$, the Gibbs law implies \cite{gibs}.
\begin{eqnarray}\label{fentropyf}
T_{in}\frac{dS_{in}}{dt}=\frac{dE}{dt}+p\frac{dV}{dt}=V\frac{d\rho}{dt}-(\rho+p)\frac{4\pi
\dot{H}}{H^4},
\end{eqnarray}
where $T_{in}\geq0$ is the temperature corresponding to the
confined fluids. Now, using~(\ref{ocon1}) and
$V=\frac{4\pi}{3H^3}$ to get
\begin{eqnarray}\label{fentropyf2}
T_{in}\frac{dS_{in}}{dt}=-\frac{4\pi(\rho+p)}{H^2}[1+\frac{\dot{H}}{H^2}],
\end{eqnarray}
telling us that, for $\rho+p\geq0$, $\frac{dS_{in}}{dt}\geq0$ is
obtainable when $1+\frac{\dot{H}}{H^2}\leq0$ which leads to
$H\leq\frac{1}{t}$. The latter means that for the perfect fluids
with state parameter $\omega$ which either meets the
$\omega\leq-1$ or $-\frac{1}{3}\leq\omega$ conditions,
$\frac{dS_{in}}{dt}\geq0$. Additionally, for a prefect fluid with
state parameter $-1\leq\omega\leq-\frac{1}{3}$, the $\rho+p\geq0$
condition is satisfied but $\frac{dS_{in}}{dt}\leq0$. Finally, for
a prefect fluid with state parameter $\omega$ which satisfies the
$-\frac{1}{3}\leq \omega$ condition the generalized second law of
thermodynamics ($\frac{dS_A}{dt}+\frac{dS_{in}}{dt}\geq0$) will be
satisfied if the $\rho+p\geq0$ and
$\rho_D+p_D\geq-\frac{\dot{H}}{4\pi}$ conditions are valid. It is
useful to mention here that $\omega=-1$ leads to
$\frac{dS_A}{dt}=0$ and $\frac{dS_{in}}{dt}=0$ meaning that the
generalized second law of thermodynamics is marginally satisfied.
Moreover, for a more general manner in which $\omega$ is not a
constant, by using the Raychaudhuri equation, we get
\begin{eqnarray}\label{rey10}
1+\frac{\dot{H}}{H^2}=1-4\pi(\rho+p)\frac{1}{H^2(1-\beta)-n\gamma
H^{2n}}.
\end{eqnarray}
Thus, $1+\frac{\dot{H}}{H^2}\leq0$ leads to
\begin{eqnarray}\label{rey10}
H^2(1-\beta)-n\gamma H^{2n}\leq4\pi(\rho+p),
\end{eqnarray}
which indicates that $\frac{dS_{in}}{dt}\geq0$. Therefore, if this
condition is valid, then the generalized second law of
thermodynamics will be satisfied.

For the $\rho+p<0$ case, it is obvious that, from
Eq.~(\ref{hentropyf2}), $\frac{dS_A}{dt}<0$. In addition, when $H$
meets the $1+\frac{\dot{H}}{H^2}\leq0$ condition,
$\frac{dS_{in}}{dt}\leq0$ and thus
$\frac{dS_A}{dt}+\frac{dS_{in}}{dt}<0$ meaning that the
generalized second law is not satisfied. Briefly, for a prefect
fluid with $\omega<-1$, the generalized second law is not
satisfied. If the Hubble parameter satisfies the
$1+\frac{\dot{H}}{H^2}>0$ condition Eq.~(\ref{fentropyf2}) leads
to $\frac{dS_{in}}{dt}\geq0$ and therefore, it is legally possible
to meet the generalized second law of thermodynamics. Using
Eq.~(\ref{rey}) to get
\begin{eqnarray}\label{rey1}
1+\frac{\dot{H}}{H^2}=1-4\pi(\rho+p)\frac{1}{H^2(1-\beta)-n\gamma
H^{2n}}.
\end{eqnarray}
Therefore, the $1+\frac{\dot{H}}{H^2}>0$ condition leads to
\begin{eqnarray}\label{rey2}
4\pi(\rho+p)<H^2(1-\beta)-n\gamma H^{2n}.
\end{eqnarray}
Finally, we can say that if this condition is valid, then
$\frac{dS_{in}}{dt}\geq0$ which may lead to satisfy the
generalized second law of thermodynamics.

For the horizon entropy of the flat FRW universe supported by a DE
candidate with unknown density profile $\rho_D$, we can use
Eqs.~(\ref{general}) and~(\ref{friedde}) to obtain
\begin{eqnarray}\label{general0}
\frac{dS_A}{dt}=\frac{8\pi^2(\rho+p)}{H^3},
\end{eqnarray}
whenever, it is easy to check that Eqs.~(\ref{fentropyf2})
and~(\ref{fff}) are also valid in this manner. Bearing
Eq.~(\ref{rey0}) in mind, $\dot{H}<0$ leads to
$\rho+p>-\rho_D-p_D$. Similarities with the previous case, in
which $\rho_D$ meets Eq.~(\ref{dden}), are obvious. In fact, in
order to achieve a more detailed resolution about the validity of
generalized second law of thermodynamics, we need to know the
dependence of either $\rho_D$ or $\rho$ to the Hubble parameter.
We should note again that the horizon component satisfies the
second law of thermodynamics by ($\frac{dS_A}{dt}\geq0$) if the
$\rho_D+p_D\geq-\frac{\dot{H}}{4\pi}$ and $\rho+p\geq0$ conditions
are met, which is in agreement with the Raychaudhuri
equation~(\ref{rey0}). Moreover, since $\frac{dS_{in}}{dt}\geq0$
is valid when $-\frac{1}{3}\leq\omega$, the generalized second law
of thermodynamics $\frac{dS_A}{dt}+\frac{dS_{in}}{dt}\geq0$ will
be available if the $-\frac{1}{3}\leq\omega$ and
$\rho_D+p_D\geq-\frac{\dot{H}}{4\pi}$ conditions are met
simultaneously. More studies on the availability of the second law
of thermodynamics and its generalization needs to know the exact
form of $\rho$.
\subsection{Interacting Case}
For this case, by using~(\ref{general2}), we get again
\begin{eqnarray}\label{lim}
\frac{dS_A}{dt}=-\frac{2\pi\dot{H}}{H^3}(1+\frac{4\pi\rho_D(1+\omega_D)}{\dot{H}}).
\end{eqnarray}
On one hand, when $\rho_D(1+\omega_D)\geq-\frac{\dot{H}}{4\pi}$,
since observationally $\dot{H}<0$ \cite{roos}, it seems that
$\frac{dS_A}{dt}\geq0$ is valid everywhere. On the other hand, by
combining Eqs.~(\ref{general12}),~(\ref{energymomentum})
and~(\ref{friedde}), once again we get
\begin{eqnarray}\label{lim1}
\frac{dS_A}{dt}=\frac{8\pi^2(\rho+p)}{H^3},
\end{eqnarray}
meaning that $\frac{dS_A}{dt}\geq0$ is valid everywhere, if
$\rho_D(1+\omega_D)\geq-\frac{\dot{H}}{4\pi}$ and $\rho+p\geq0$
are satisfied simultaneously. Therefore, the quality of validity
of the second law of thermodynamics is similar with the
non-interacting case. In addition, Eq.~(\ref{fentropyf}) leads to
\begin{eqnarray}\label{fentropyf22}
T_{in}\frac{dS_{in}}{dt}=-\frac{4\pi(\rho+p)}{H^2}[1+\frac{\dot{H}}{H^2}]
-V\dot{H}[\rho_D^{\prime}+3\frac{H}{\dot{H}}(\rho_D+p_D)],
\end{eqnarray}
where prime denotes derivative with respect to the Hubble
parameter, again. Here, we focus on the model proposed by Lime st
al. \cite{lima}. In this model, $\omega_D=-1$ while the vacuum
density meets Eq.~(\ref{dden}). Substituting into the above
equation to get
\begin{eqnarray}\label{fentropyf222}
T_{in}\frac{dS_{in}}{dt}=-\frac{4\pi(\rho+p)}{H^2}[1+\frac{\dot{H}}{H^2}]
-V\dot{H}[\frac{3\beta H+ 3n\gamma H^{2n-1}}{4\pi}].
\end{eqnarray}
Since $\dot{H}<0$, the second term of RHS of this equation
($-V\dot{H}\rho_D^{\prime}$) is positive everywhere and therefore,
the validity of $\frac{dS_{in}}{dt}>0$ and thus the generalized
second law of thermodynamics depends on the value of the first
term of RHS ($-\frac{4\pi(\rho+p)}{H^2}[1+\frac{\dot{H}}{H^2}]$).
It is useful to mention here that for a perfect fluid either
obeying $\omega\leq-1$ or $-\frac{1}{3}\leq\omega$, the Hubble
parameter meets the $H\leq\frac{1}{t}$ condition leading to
$1+\frac{\dot{H}}{H^2}\leq0$ and thus $\frac{dS_{in}}{dt}>0$.
Moreover, from Eqs.~(\ref{lim}) and~(\ref{lim1}) it is apparent
that $\frac{dS_{A}}{dt}>0$ when $\omega_D=-1$ and $-1\leq\omega$,
respectively. Therefore, for the flat FRW universe embraced a
prefect fluid which satisfies $-\frac{1}{3}\leq\omega$, the
generalized second law of thermodynamics is convinced. As again,
more studies on the availability of the second law of
thermodynamics and its generalization needs to know the exact form
of $\rho$.
\section{Summary and concluding remarks}
Throughout this paper, we considered the FRW universe filled by a
DE candidate together a fluid, which is the agent of the other
possible sources, which may include the baryonic and non-baryonic
matters, enclosed by the apparent horizon of the flat FRW
universe. In continue, we proposed a profile density for the DE
candidate which covers proposals including CC and dynamic models
of DE such as ghost dark energy model, its generalization, the
Lima's model and etc. Moreover, by taking into account the
Hayward-Kodama definition of the temperature definition of
apparent horizon as well as the Friedmann equation, we could find
the horizon entropy for models in which the DE candidate does not
interact with the other parts of the cosmos. Our study shows that
the DE candidate may modify the horizon entropy. We have shown
that our formula for entropy~(\ref{hentropy}) is compatible with
previous results about the ghost dark energy and its
generalization \cite{cana,cana1}. Indeed, similar result
with~(\ref{hentropyggde}) is reported by authors in
ref.~\cite{cana}. But, our derivation is completely different.
Here, we have considered the apparent horizon as the causal bound
of the system, whiles authors in~\cite{cana} used the trapping
horizon as the causal bound to get the associated horizon entropy.
In addition, we have generalized our formulation to models in
which the DE candidate is an arbitrary unknown function, and
showed that the DE candidate may modify the horizon
entropy~(\ref{general1}) independent of the other parts of cosmos.
We have also used the Cai-Kim temperature to get the horizon
entropy, and found out that the same result for the horizon
entropy is obtainable if one considers an infinitesimal time in
which $dV=0$. Thereinafter, we focused on the models in which the
DE candidate interacts with the other parts of cosmos. We found
that the mutual interaction between the cosmos sectors may also
modify the apparent horizon entropy~(\ref{general12}). Our study
shows that for models in which $\omega_D=-1$, such as the model
proposed by Lima et al.~\cite{lima}, the mutual interaction
between the cosmos sectors does not disturb the Bekenstein limit
of the horizon entropy. It means that there is no modification to
the horizon entropy for interacting models with $\omega_D=-1$ and
therefore, $S_A=\frac{A}{4}$ is available in these models. The
same as the non-interacting case, we tried to get a relation for
the horizon entropy in the interacting models by using the Cai-Kim
temperature. Our study shows that the same result as that of
obtained by considering the Hayward-Kodama temperature is
available for the horizon entropy. Additionally, we pointed to the
some required conditions for availability of the second law of
thermodynamics and its generalization in the interacting and
non-interacting models. Our studies show that for the
non-interacting case, whiles
$\rho_D+p_D\geq-\frac{\dot{H}}{4\pi}$, the second law of
thermodynamics and its generalization are inevitably valid if the
state parameter of other parts of the cosmos satisfies the
$-\frac{1}{3}\leq\omega$ condition. It is because
$\frac{dS_{A}}{dt}>0$ and $\frac{dS_{in}}{dt}>0$ are separately
valid in this situation. Finally, our study shows that for the
interacting case with $\omega_D=-1$, $\frac{dS_{A}}{dt}>0$ and
$\frac{dS_{in}}{dt}>0$ will be met if $-\frac{1}{3}\leq\omega$ and
therefore, the generalized second law of thermodynamics will be
available in an unavoidable way.
\section*{Acknowledgments}
We are grateful to the anonymous referee for the constructive
worthy comments which help us increase our understanding of the
subject. The work of H. M. has been supported financially by
Research Institute for Astronomy \& Astrophysics of Maragha
(RIAAM) under research project No. $1/4165-6$.

\end{document}